\documentclass[a4paper,11pt]{article}

\usepackage{lineno}
\usepackage{xcolor}
\usepackage{booktabs}
\usepackage{longtable}
\usepackage{algpseudocode}
\usepackage[normalem]{ulem}

\usepackage{graphicx}
\usepackage{subcaption}
\graphicspath{ {./figures/} }

\usepackage{jheppub} 

\usepackage{amsmath}
\usepackage{amsthm}
\usepackage{mathtools}

\title{Holographic Entanglement Distillation from the Surface State Correspondence}

\author[a,b]{Ning Bao,}
\author[a]{Gün Süer}
\affiliation[a]{Department of Physics, Northeastern University, Boston, MA 02115, USA}
\affiliation[b]{Computational Science Initiative, Brookhaven National Laboratory, Upton, NY 11973, USA}

\emailAdd{ningbao75@gmail.com}
\emailAdd{gunsuer06@gmail.com}

\abstract{We study correlations between geometric subfactors living on the Ryu-Takayanagi surface that bounds the entanglement wedge. Using the surface-state correspondence and the bit threads program, we are able to calculate mutual information and conditional mutual information between subfactors. This enables us to count the shared Bell pairs between subfactors, and we propose an entanglement distillation procedure over these subsystems via a SWAP gate protocol. We comment on extending to multipartite entanglement.}

\makeatletter\def\@fpheader{~}\makeatother

\begin{document}
\maketitle
\flushbottom

\section{Introduction}
In recent years there has been great progress made at the intersection of quantum information and quantum gravity. One of the major areas of interest has been the study of entanglement entropies in holographic settings \cite{Ryu_2006}. In this context, the Ryu-Takayanagi (RT) formula relates the  entanglement entropy of a boundary subregion $A$ to the area of the minimal surface $\gamma$ in the bulk homologous to that region
\begin{equation}
    S(A) = \min_{\partial\gamma = \partial A} \frac{\text{Area}(\gamma)}{4G_N}.
\end{equation}
In addition to its conceptual appeal in generalizing the area law for black hole entropy to entanglement entropies in conformal field theories, it also enables purely geometrical proofs of various information-theoretic relations such as strong subadditivity \cite{Headrick_2007} and higher party number generalizations of entanglement entropy inequalities for holographic states \cite{Bao_2015}.

Since entanglement is a central resource of quantum information processing systems, the question of available entanglement in a given system is of clear importance \cite{Chitambar_2019}. Given an arbitrarily entangled quantum state, the process of transforming the initial state to a certain number of Bell pairs using local operations and classical communication (LOCC) is known as entanglement distillation \cite{Deutsch:1996um}. While algorithms for performing entanglement distillation for pure states are known, for generic mixed states even the amount of distillable entanglement is not known in general.

In this direction, a more recent entry in the holographic dictionary is the $E_W = E_P$ conjecture \cite{Nguyen_2018,Umemoto_2018}. It states that the entanglement wedge cross-section $E_W$ separating two regions is equal to the entanglement of purification $E_P$, on the basis that the holographic object $E_W$ obeys the same set of inequalities as the information-theoretic object $E_P$. Given a bipartite state $\rho_{AB}$, the entanglement of purification is defined as
\begin{equation}
    E_P(A:B) = \min \left\{S(AA')\ | \ \rho_{AA'BB'} \ \text{is pure}\right\},
\end{equation}
which measures correlations in a bipartite mixed state. Since it involves optimizing over the purification, it is very challenging to compute. Therefore finding a holographic dual to entanglement of purification and calculating it using geometrical tools is appealing. It has been conjectured that the entanglement of purification is exactly equal to the area of the entanglement wedge cross-section which is the minimal surface that splits the entanglement wedge $r_{AB}$ into two regions
\begin{equation}
    E_W(A:B) = \min \left\{\text{Area}(\Gamma_{AB})\ | \ \Gamma_{AB}\subset r_{AB}\text{ splits }r_{AB}\text{ into two regions}\right\}.
\end{equation}

In this paper we focus on understanding entanglement distillation in a holographic setting, using the surface state correspondence \cite{Miyaji_2015} and the bit threads program \cite{Freedman_2016}. Specifically, we consider two boundary region entanglement wedges and calculate mutual information and conditional mutual information for subfactors living on RT surfaces.\footnote{From the bit threads program \cite{Freedman_2016} one can see that the RT surface corresponding to a given subregion acts as a geometric bottleneck for the number of bit threads through the area. In conjunction with the surface state correspondence, this leads to the fact that the number of qubits living on the RT surface is precisely given by the RT surface area. Therefore, assuming the surface state correspondence, one can use the terms subregion and subfactor interchangeably on such surfaces. We apologize for our abuse of notation and use these qualities interchangeably for the rest of the paper.} The states on the surfaces in the bulk are related to states on the boundary via an isometry, which is a local unitary preserving the entanglement properties. By selecting pairs of qubits across different subsystems and performing SWAP tests, we present a straightforward entanglement distillation procedure that works at least asymptotically. A direction complementary to our line of work focuses on the entanglement structure of surface states in the context of tensor networks \cite{Akers:2023obn, Mori_2022}.

\begin{figure}
    \centering
    \includegraphics{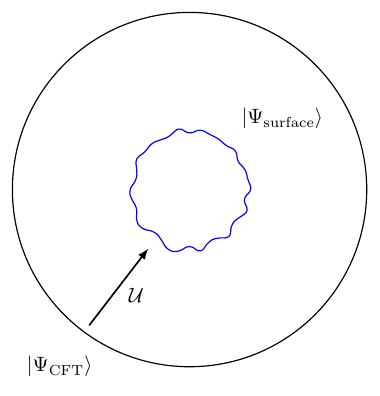}
    \caption{Under the surface-state correspondence, the boundary can be pushed to a closed surface in the bulk via an isometry. The isometry used is a local unitary acting only on the subregion being pushed, preventing the change of entanglement properties.}
    \label{fig:boundary}
\end{figure}

\section{Surface State Correspondence}
\begin{figure}
    \centering
    \includegraphics{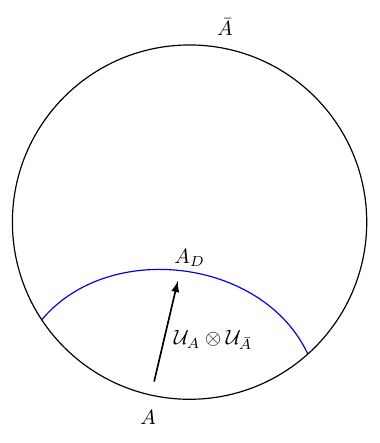}
    \caption{One can also consider only pushing a subregion while keeping its complement fixed. This preserves the entanglement properties across the bipartition.}
    \label{fig:1-party}
\end{figure}

The surface state correspondence is a conjectured property of the AdS/CFT correspondence. It states that one can define a pure state on any closed surface not containing topological obstructions in the bulk. This surface is related to the boundary by continuous deformation, or pushing inwards from the boundary into the bulk. Such pushing, again in the absence of topological obstructions, will always result in a closed surface; see Fig. \ref{fig:boundary}. The boundary state is then related to the state on this closed bulk surface via isometry. In particular, we are allowed to push only a geometric subregion of the boundary, while keeping the remainder of the boundary fixed; see Fig. \ref{fig:1-party}. At the conclusion of this pushing, the new closed surface functions as the new boundary, for the purposes of homology constraints of the RT formula. Once this pushing has been performed, a new boundary subregion of this redefined boundary, here taken to not overlap with the first boundary subregion selected, can be chosen for pushing.\footnote{The non-overlapping condition of this pushing is a special instance of the tree tensor network style pushing of \cite{Bao_20191}.}

There will be a portion of the state which is discarded in the process of implementing the isometry, e.g. a portion of the state which was defined on a subfactor of the original boundary Hilbert space but is not supported on the Hilbert space of the closed curve. This portion of the state will be associated with the region that the isometry pushed through, and in particular will not be entangled with either the surface state or the remainder of the boundary CFT state, due to the purity of the surface state and the fact that the isometry used was a local unitary acting only on the subregion being pushed and not its complement, preventing the change of entanglement properties across the bipartition.

In this work, we will focus on pushing from boundary subregions to their RT surfaces. By purity of the overall closed surface state, this pushing must preserve the entanglement entropy of the boundary region. However, because the RT surface has an area precisely equal to the entanglement entropy of the boundary region up to normalization, by the bottleneck argument in the equivalent bit threads formalism \cite{Freedman_2016} the log of the Hilbert space dimension of the portion of the state associated with a RT surface is equal to its entanglement entropy. This means that the portion of the state living on an RT surface in the surface state correspondence must be maximally mixed.

We will leverage this argument to derive several strong results on entanglement distillation for mixed states in holography.

\section{Two Boundary Region Entanglementment Wedges}
Let's start with a connected entanglement wedge bounded by two disconnected boundary regions, $A$ and $B$. We restrict our attention to this case because the case of the disconnected entanglement wedge is trivial for our considerations. We will first push the complementary boundary region to the RT surface of $\rho_{AB}$, as for pure CFT states the RT surface of the complement is the same as that of $AB$. Because surface-state implies the $E_P=E_W$ conjecture \cite{Umemoto_2018}, we can associate the $A'$ factor of the purification to the portion of the RT surface on the same side of the $E_W$ surface as $A$, and the $B'$ factor of the purification to the portion of the RT surface on the same side of the $E_W$ surface as $B$. Note that by the arguments of the previous section, $\rho_{A'B'}$ must be maximally mixed, to leading order in $1/N$.

\begin{figure}
    \centering
    \includegraphics{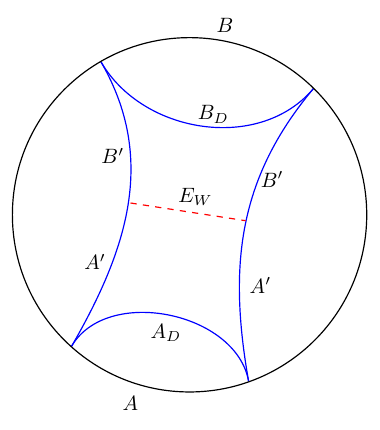}
    \caption{We push the regions $A$ and $B$ to their respective RT surfaces, $A_D$ and $B_D$. $A'B'$ is the entanglement wedge. $E_W$ is the minimal entanglement wedge cross-section. It bounds the regions $A'$ and $B'$ with $A$ and $B$ respectively.}
    \label{fig:2-party}
\end{figure}

At this point, we will use the iterative step from the previous section to go further. Keeping the $A'B'$ portion of the surface fixed, we can now also push $\rho_A$ and $\rho_B$ to their respective RT surfaces; see Fig. \ref{fig:2-party}. We will call the states here $\rho_{A_D}$ and $\rho_{B_D}$. Similarly to above, $\rho_{A_D}$ and $\rho_{B_D}$ must also be maximally mixed, to leading order in $1/N$.

\subsection{Mutual Information}
Now we can compute mutual information between pairs of these regions. First, and most simply, $I(A':B')=0$, as the mutual information between two subfactors of a maximally mixed state is always zero. 

Now we turn our attention to $I(A:B)$. Because only the $A_D$ and $B_D$ subfactors can contribute to this mutual information, we have that

\begin{equation}
    I(A:B)=I(A_D:B_D)=S(A_D)+S(B_D)-S(A_D B_D)=S(A)+S(B)-S(AB).
\end{equation}
This might seem trivial, but it's important to note that due to $\rho_{A_D}$ and $\rho_{B_D}$ being maximally mixed, their entanglement entropies are the logs of their Hilbert space dimensions. Moreover, because they are maximally mixed, the only internal purification in $\rho_{A_D B_D}$ allowed are Bell pairs formed between qubits in $A_D$ and qubits in $B_D$. Therefore, this mutual information counts the number of Bell pairs between qubits in $A_D$ and those in $B_D$, as it sums the individual entropies which count the number of qubits in $A_D$ and $B_D$, and subtracts off those that are not internally purified. Because the mutual information is an upper bound on the distillable entanglement between $A_D$ and $B_D$, this is sufficient to show that to leading order the holographic distillable entanglement is equal to $I(A_D:B_D)=I(A:B)$, as these Bell pairs between $A_D$ and $B_D$ are certainly distillable across the bipartition.

We can also compute $I(A:A')$. By the same logic before,

\begin{equation}
    I(A:A')=S(A_D)+S(A')-S(AA').
\end{equation}
Here we mix the $A_D$ and $A$ Hilbert space factors for convenience. The subtracted term is clearly the $E_P$ quantity.\footnote{Again, we are using the fact that surface state implies $E_P=E_W$.} But, by the same arguments in considering $I(A:B)$, this quantity computes the amount of distillable entanglement between $A$ and $A'$.

Finally, we come to $I(A:B')$. By the same arguments as above, this computes the amount of distillable entanglement between $A$ and $B'$. However, the expanded equation merits further study:

\begin{equation}
    I(A:B')=S(A_D)+S(B')-S(AB').
\end{equation}
The first two terms are simply the area of the RT surface of $A$ and the $B'$ portion of the RT surface for $B'$, respectively. However, the subtracted term, $S(AB')=\min (S(A)+S(B'),S(A')+S(B))$, as these are the two candidate RT surfaces within the surface state region. The former term would return zero, which means that wlog either $I(A:B')=0$ or $I(B:A')=0$, for all holographic states at leading order.\footnote{For symmetric cases both terms in the minimization would be zero, but generically one of them will be nonzero.} This would also mean that the distillable entanglement between these systems is also zero, as again the mutual information upper bounds the distillable entanglement. In the nonzero instance, $I(A:B')=S(A)-S(A')-(S(B)-S(B'))$.

\subsection{Entanglement Distillation Procedure}
While for generic mixed states entanglement distillation procedures are not known, for this holographic case it appears quite straightforward, at least asymptotically. We describe the procedure for $\rho_{AB}$, but it works the same way for the other choices of subsystems. Select a qubit in $A_D$ and a qubit in $B_D$ and consider this two-qubit state, while also selecting two other qubits, both in $A_D$, and consider its two-qubit state. The former can either be a Bell pair or a maximally mixed state, and the latter can only be a maximally mixed state.\footnote{This is true up to subleading corrections, so a more precise statement would be that the trace distance between this state and the maximally mixed state is zero up to $1/N$ corrections.} Now simply perform the SWAP test between these two two-qubit states. With enough copies of the CFT, one can repeat this for all possible two-qubit states with one qubit from $A_D$ and one from $B_D$, and find all of the Bell pairs in worst case $O(N_{A_D} N_{B_D})$ trials, and from there on simply read off the Bell pairs from the remaining infinite number of copies of states, asymptotically achieving $I(A:B)$.

\subsection{Conditional Mutual Information}
We can also compute conditional mutual information using the same technique, and we describe these calculations below, with minimal comment.

There are twelve distinct conditional mutual informations: $I(A:B|A')$, $I(A:B|B')$, $I(A:B'|B)$, $I(A:B'|A')$, $I(A:A'|B)$, $I(A:A'|B')$, $I(A':B|A)$, $I(A':B|B')$, $I(A':B':A)$, $I(A':B':B)$, $I(B:B'|A)$, and $I(B:B'|A')$. Wlog, we take $S(A'B)=S(A')+S(B)=S(AB')$.

In order,
\begin{align}
    I(A:B|A')=S(AA')+S(A'B)-S(A')-S(B')\\
    =E_W(A:B)+S(B)-S(B'),
\end{align}

\begin{align}
    I(A:B|B')=S(AB')+S(B'B)-S(A')-S(B')\\
    =E_W(A:B)+S(B)-S(B').
\end{align}
The first two are the same.

\begin{align}
    I(A:B'|B)=S(AB)+S(BB')-S(B)-S(A')\\
    =E_W(A:B)+S(B')-S(B),
\end{align}

\begin{align}
    I(A:B'|A')=S(AA')+S(A'B')-S(B)-S(A')\\
    =E_W(A:B)+S(B')-S(B).
\end{align}
These two are also the same.

\begin{align}
    I(A:A'|B)=S(AB)+S(A'B)-S(B)-S(B')\\
    =S(AB)+S(A')-S(B')\\
    =2S(A').
\end{align}

\begin{align}
    I(A:A'|B')=S(AB')+S(A'B')-S(B')-S(B)\\
    =S(AB)+S(A')-S(B')-S(B)\\
    =2S(A').
\end{align}
These two are the same.

\begin{align}
    I(A':B|A)=S(AA')+S(AB)-S(A)-S(B')\\
    =E_W(A:B)+S(A')-S(A),
\end{align}

\begin{align}
    I(A':B|B')=S(A'B')+S(BB')-S(A)-S(B')\\
    =E_W(A:B)+S(A')-S(A).
\end{align}
These two are the same.

\begin{align}
    I(A':B'|A)=S(AA')+S(AB')-S(A)-S(B)\\
    =E_W(A:B)+S(A')-S(A),
\end{align}

\begin{align}
    I(A':B'|B)=S(BB')+S(AB')-S(A)-S(B)\\
    =E_W(A:B)+S(A')-S(A).
\end{align}
These two are the same.

\begin{align}
    I(B:B'|A)=S(AB)+S(AB')-S(A)-S(A')\\
    =2S(A')+S(B')+S(B)-S(A),
\end{align}

\begin{align}
    I(B:B'|A')=S(AB)+S(AB')-S(A)-S(A')\\
    =2S(A')+S(B')+S(B)-S(A).
\end{align}
These two are the same.

It is worth noting that all of these quantities must be non-negative, and moreover must lower bound the mutual information, thereby giving some lower bounds for the distillable entanglement.
\subsection{Tripartite Information}
We can also compute tripartite information. There are four tripartite informations to keep track of: $I(A:B:A')$, $I(A:B:B')$, $I(A:A':B')$, and $I(B:A':B')$. Wlog, we are taking $S(A'B)=S(A')+S(B)=S(AB')$.

In order,
\begin{align}
    I(A:B:A')=S(AB)+S(AA')+S(A'B)-S(A)-S(B)-S(A')-S(B')\\   
    =S(AA')+S(A'B)-S(A)-S(B)\\
    =E_W(A:B)+S(A')-S(A).
\end{align}

\begin{align}
    I(A:B:B')=S(AB)+S(AB')+S(BB')-S(A)-S(B)-S(A')-S(B')\\
    =S(BB')+S(AB')-S(A)-S(B)\\
    =E_W(A:B)+S(A')-S(A).
\end{align}

\begin{align}
    I(A:A':B')=S(AA')+S(AB')+S(A'B')-S(A)-S(A')-S(B')-S(B)\\
    =S(AA')+S(AB')-S(A)-S(B)\\
    =E_W(A:B)+S(A')-S(A).
\end{align}

\begin{align}
    I(B:A':B')=S(BA')+S(BB')+S(A'B')-S(A)-S(A')-S(B')-S(B)\\
    =S(BB')+S(AB')-S(A)-S(B)\\
    =E_W(A:B)+S(A')-S(A).
\end{align}
In the end, all four tripartite contributions are the same as the Markov gap noted in \cite{Hayden_2021}, as expected.
\section{Three Boundary Region Entanglement Wedges}
We can extend the analysis from before to three boundary regions, $A$, $B$, and $C$, with $A'$, $B'$, and $C'$ defined as in \cite{bao2018holographic}. We will again do the pushing to the RT surface of $ABC$, $A$, $B$, and $C$ in the analogous way.
\subsection{Mutual Information}
We can now compute mutual information in analogy to before. The mutual information of the primed regions remains zero, as before: $I(A':B'C')=0$. The mutual information $I(A:BC)$ is the same and still computes the distillable entanglement between $A$ and $BC$.

Differences arise when we consider $I(A:B')$,
\begin{equation}
I(A:B')=S(A)+S(B')-\min(S(A)+S(B'), S(B)+S(A')+S(CC')).    
\end{equation} 
So, if Wlog $I(A:B')\neq 0$, this equals $S(A)-S(A')-(S(B)-S(B'))-S(CC')$, which again must be nonnegative by dint of being a mutual information, providing a nontrivial bound. Note that this does not become zero in the symmetric case, as with the two boundary situations.

There are too many different entropic combinations to consider them all for the three-party case, but clearly, there is a rich structure here that is worthy of further exploration.

\section{Conclusion and Outlook}
In this work, we have investigated the bipartite and tripartite entanglement properties of subfactors on RT surfaces inspired by the recently proposed $E_W = E_P$ conjecture. By using the surface state correspondence and the bit threads program we calculate mutual information and conditional mutual information across different subfactors living on the boundary of the entanglement wedge. After estimating the number of shared Bell pairs across different regions, we employ a SWAP test for entanglement distillation, using the fact that the states on RT surfaces in the surface state correspondence are maximally mixed. We leave entanglement properties and distillation in multipartite systems for future work.

\acknowledgments
We are grateful to Charles Cao and Jonathan Harper for discussions. N.B.
is supported by the Computational Science Initiative at Brookhaven National Laboratory, Northeastern University, and by the U.S. Department of Energy QuantISED Quantum Telescope award. G.S. is supported by the Graduate Assistantship from the
Department of Physics, Northeastern University.


\bibliographystyle{JHEP}
\bibliography{biblio.bib}

\end{document}